%% file: evolpisn7.tex
\documentclass[structabstract]{aa}  
\usepackage{graphicx}
\usepackage{txfonts}
\usepackage{natbib} 
\usepackage{url} 
\usepackage{subfig} 
\usepackage{multirow} 
\newcommand{\Msun}{$M_\odot$}

\newcommand{\mb}[1]{\ensuremath{\mathbf{#1}}}
\newcommand{\mr}[1]{\ensuremath{\mathrm{#1}}}
\newcommand{\pd}[2]{\frac{\partial #1}{\partial #2}}

\begin{document}
\title{Explosion and nucleosynthesis of \\ low redshift
pair instability supernovae}

\author{A.~Kozyreva \inst{1}
\and
S.-C.~Yoon \inst{1,2}
\and
N.~Langer \inst{1}
}

\institute{
Argelander-Institut f\"ur Astronomie, Universit\"{a}t Bonn, Auf dem H\"ugel 71, 53121 Bonn, Germany \\
\email{kozyreva@astro.uni-bonn.de}
\and
Astronomy Program, Department of Physics \& Astronomy, Seoul National University, Seoul, 151-742, Republic of Korea
             }
   \date{Received XXXXX XX, 2014; accepted XXXXX XX, 2014}

 
  \abstract
{ 
Both recent observations and stellar evolution models suggest that pair-instability supernovae (PISNe)
could occur in the local Universe, at metallicities below $\lesssim Z_\odot/3${}.  Previous PISN models were mostly
produced at very low metallicities in the context of the early Universe.
}
{We present new PISNe models at a metallicity of {\emph Z} = 0.001, which are relevant for the local Universe.
}
{
We take the self-consistent stellar evolutionary models of pair-instability progenitors with initial masses of 150~\Msun{} and 
250~\Msun{} at metallicity of {\emph Z} = 0.001 by Langer~et~al.~(2007) and follow the 
evolution of these models through the supernova explosions, 
using a hydrodynamics stellar evolution code with an extensive nuclear network including 200~isotopes. 
}
{
Both models explode as PISNe without leaving a compact stellar remnant.  
Our models produce a nucleosynthetic pattern that is generally similar to that
of Population~III PISN models, which is mainly characterized by the production
of large amounts of $\alpha-$elements and a strong deficiency of the
odd-charged elements.  However, the odd-even effect in our models is significantly weaker than that found in Population~III
models.  The comparison with the nucleosynthetic yields from
core-collapse supernovae at a similar metallicity ($Z = 0.002$) indicates that
PISNe could have strongly influenced the chemical evolution below
$Z \approx 0.002$, assuming a standard initial mass function.  
The odd-even effect is predicted to be most prominent for the
intermediate mass elements between silicon and calcium.  
}
{
With future
observations of chemical abundances in Population~II stars, our result can
be used to constrain the number of PISNe that occurred during the past evolution of our Galaxy.
}


   \maketitle
%

\section{Introduction}
\label{sect:intro1}

The pair instability mechanism for supernova explosions was first
suggested in 1960s
\citep{1964ApJS....9..201F,1967SvA....10..604B,1967ApJ...148..803R,1967PhRvL..18..379B,1968Ap&SS...2...96F,1971reas.book.....Z}.
The cores of very massive stars with initial masses higher than about
100~$M_\odot${} \citep{1982sscr.conf..303B,2003ApJ...591..288H} have relatively
low densities and high temperatures for which radiation pressure is dominant
over gas pressure.  When the core temperature approaches $10^{\,9}$\,K, the creation
of electron-positron pairs out of gamma-ray photons from the high energy tail
of the black body spectrum becomes important and makes the adiabatic index $\Gamma$
drop below 4/3.  This causes gravitational collapse of the core if a
significant fraction of the core has $\Gamma < 4/3$.  
The consequent oxygen burning induces a thermonuclear explosion that
completely disrupts the star if the released energy exceeds its binding energy.  
This happens for oxygen core masses above approximately 45~$M_\odot${}.

For a pair instability supernova to occur, its progenitor needs to retain its mass high
enough to keep its helium core mass above about $\sim\,65~M_\odot${}.
This condition cannot be easily fulfilled
at high metallicity for which the evolution of very massive stars are dominated
by stellar wind mass-loss \citep[e.g.][]{2011A&A...531A.132V}.  
This is the reason why most theoretical
studies of pair instability supernovae (PISNe) have focused on zero or extremely metal poor stars in
the early Universe \citep{1983A&A...119...54E,2002ApJ...565..385U,2002ApJ...567..532H,
2005ApJ...633.1031S,2011ApJ...734..102K,2012MNRAS.422.2701P,2013ApJ...777..110W,
2013MNRAS.428.3227D}.  However, \citet{2007A&A...475L..19L} recently pointed
out that the metallicity threshold for PISNe can be as high as $Z_\odot/3${}
within the current theoretical uncertainty of stellar wind mass-loss rates,
implying one PISN per one thousand supernovae in the local Universe. 

PISNe would be marked by broad light curves given their high progenitor masses.
They would also appear extremely luminous if their progenitors have large radii
and/or if a large amount of nickel is produced as a result of the pair creation instability
\citep[e.g.,][]{2005ApJ...633.1031S,2011ApJ...734..102K}.   This raises the question whether some
of the  super-luminous SNe of various types like SN 2006gy and SN 2007bi
discovered in the nearby Universe have a pair instability origin \citep[see][for a
review]{2012Sci...337..927G}.  For example, the light curve of SN~2007bi
implies the radioactive decay of more than 3~\Msun{} of nickel, for which a pair instability
explosion gives one of the best explanations \citep{2009Natur.462..624G}.
Alternative possibilities are supernova powered by a young magnetar as
suggested by various authors \citep{2010ApJ...717..245K,2012MNRAS.426L..76D} 
and interaction-driven supernova \citep{2010ApJ...717L..83M}.  
If local PISNe would exist, one would have to wonder how they would have impacted
on the chemical evolution of the local Universe.

Addressing these questions requires PISN models that are relevant to the
environment of the local Universe.  
The first studies of local PISN models were performed by
\citet{1986A&A...167..265L}, \citet{1986A&A...167..274E} and
\citet{1990A&A...233..462H} who calculated evolutionary models with an initial
mass of 100~\Msun{} and a metallicity of $Z = 0.03$.  More recently,
\citet{2007A&A...475L..19L} calculated 150~$M_\odot${} and 250~$M_\odot${} models
at a metallicity of $Z = 0.001$ as PISN progenitors, adopting the most up-to-date prescriptions for
stellar wind mass-loss rates.  These models provide
self-consistent progenitor models for PISNe in the local
Universe together with more recent models by \citet{2013MNRAS.433.1114Y}.  
In the present study we follow the
evolution of two models from \citet{2007A&A...475L..19L} through the explosive oxygen and silicon burning stages to
verify that they explode via the pair instability mechanism, and to discuss implications
for nucleosynthesis in the local Universe. Their shock-breakout
signatures and light curves will be discussed  in a separate paper \citep{2014arXiv1403.5212K}.

This paper is organized as follows.  We describe the numerical  method adopted 
in the present study  in Section~\ref{sect:method1}.  The results of our calculations
are reported in Section~\ref{sect:result1}, where we also discuss the nucleosynthesis yields of our
PISN models.  We discuss the implications of our results for the chemical
evolution of the local Universe in Section~\ref{sect:discussion1}, and conclude our
study in the final section. 

\section{Numerical method and input physics} 
\label{sect:method1}

We use an implicit Lagrangian hydrodynamics code which solves the difference
equations for the stellar structure iteratively by the Henyey relaxation method
\citep{1964ApJ...139..306H,2000ApJ...528..368H,2005A&A...443..643Y,2006A&A...460..199Y}.  We list the relevant stellar structure
equations in the Appendix~\ref{appendix:append}.  The
equation of state is based on \citet{1996ApJS..106..171B} and includes ions,
electrons and positrons, radiation, degeneracy effects and ionization
contributions.  The opacity is computed from the OPAL tables
\citep{1996ApJ...464..943I} and \citet{1994ApJ...437..879A}.

We compute the nucleosynthesis and the corresponding energy generation rate in the
following way.  For temperatures less than $4.5\times 10^{\,8}$~K, a small nuclear
network (39 isotopes) is utilized.  For higher temperatures we use the ``Torch''
nuclear network developed by
\citet{Timmes..public..networks,1999ApJS..124..241T} with 200 isotopes.  In
this network, the weak interactions are followed using the data provided by
\citet{1982ApJS...48..279F}.  For a temperature range where silicon burning is
well described in terms of quasi-statistical equilibrium 
\citep[QSE,][]{1968ApJS...16..299B,1996ApJ...460..869H}, energy generation rates can be
given as a function of temperature $T$, density $\rho$, total mass fraction of
the silicon QSE-group\footnote{QSE-groups are the groups of isotopes formed in the
condition of quasi-statistical equilibrium \citep{1997RvMP...69..995W}.}
elements $X_{\mathrm{Si}}$ and electron abundance $Y_e$.
For calculating energy generation rates during silicon burning, therefore, we
use an energy generation rate table for a number of combinations of different
physical parameters ($T = (2.4-5) \times 10^{\,9}$~K, 
$\log_{10}(\rho~/~(\mathrm{g}~\mathrm{cm}^{-3})) = 5-10$, $X_{\mathrm{Si}} = 0~-~1$, $Y_e~=~0.44~-~0.5$), 
following \citet{Nomoto198813}.  For very high temperatures ($T~>~5~\times~10^{\,9}$~K), 
the nuclear statistical equilibrium routine by \citet{Timmes..public..networks} is
employed.

Our starting models  are taken from the stellar evolutionary calculations with an initial
masses of 150~\Msun{} and 250~\Msun{} and initial rotational velocity of
$10~\mathrm{km~s^{-1}}$ at  $Z=0.001$ by \citet[][the model Sequences 3 and 4
]{2007A&A...475L..19L}.  These models were calculated from
the zero-age main sequence  until the onset of the pair instability in the core, with the
stellar wind mass-loss prescription described in \citet{2006A&A...460..199Y}.  
The starting point of our calculations is core carbon exhaustion, after which
these stars quickly enter the pair instability phase. The stellar masses at this point are
94~\Msun{} and 169~\Msun{} for the 150~\Msun{} and 250~\Msun{} stars, respectively.
We summarize some model properties in Table~\ref{table:modeldata} along with
those of zero metallicity models by \citet{2002ApJ...567..532H} for comparison.

The stellar evolutionary models from \citet{2007A&A...475L..19L} were calculated using the Ledoux criterion for convection,
with the assumption of semi-convection \citep{1983A&A...126..207L} using a large semi-convective mixing
parameter ($\alpha_{\mathrm{SEM}}=1$, \citet{1991A&A...252..669L}), and without convective core overshooting.  The mixing length parameter
was chosen to be 1.5 of pressure scale height \citep{2006A&A...460..199Y}.  
We neglected the convective mixing during the explosive oxygen and silicon burning phases in
our calculations because the convection timescale is two orders of magnitude
larger than the hydrodynamical timescale on which the collapse induced by the
pair instability develops.  

Note that the recent PISN progenitor models at higher metallicity (0.002 and 0.006) by \citet{2013MNRAS.433.1114Y}
are computed using the Schwarzschild criterion for convection and with core overshooting with a moderate overshooting parameter
($\alpha_{\mathrm{over}}=0.1$).  The convection in the outer layers is calculated with mixing-length parameter scaled to
the density scale height ($\alpha_{\mathrm{MLT}}= l / H_\rho = 1$) to avoid density inversions \citep[see
also][]{2012A&A...537A.146E}.  The consequences of this treatment are more compact stellar models, a lesser degree of mass loss,
and larger carbon-oxygen cores.  
Those models from \citet{2013MNRAS.433.1114Y} which are supposed/declared to produce PISNe are evolved until the end of helium/oxygen
burning.  Electron-positron pair creation is not included in the equation of state of the employed evolutionary code.  The statement about
the PISN fate is based on the size of the carbon-oxygen core.  In more recent study the PISN models from
\citet{2013MNRAS.433.1114Y} (at the end of core helium
burning) were mapped into the KEPLER code \citep[][]{2010ApJ...724..341H}.  
With these calculations the models were evolved through pair instability and eventually
exploded \citep{2013arXiv1312.5360W}.

Rotation is not included during the present calculations because
these models retain very small amounts of angular momentum. 

\begin{table*}
\caption[Properties of our PISN progenitor models and of comparable Population~III helium star models]
{Properties of our PISN progenitor models and of comparable 
Population~III helium star models from \citet{2002ApJ...567..532H}. 
$T_{\mathrm{c}}^{\,\mathrm{max}}$ and $\rho_{\mathrm{c}}^{\,\mathrm{max}}$ are the maximum values of 
central temperature and central density, respectively, that
are achieved during the calculations.  The last two columns
give the values of the central neutron excess initially and at maximum temperature.
}\label{table:modeldata}
\begin{center}
\begin{tabular}{rccccccc}
\hline
Initial mass  & Final mass   & He-core & O-core & $T_\mathrm{c}^{\,\mathrm{\max}}$ & $\log \rho_\mathrm{c}^{\,\mathrm{\max}}$ & $\eta_\mathrm{c}^{\,\mathrm{init}}$ & $\eta_\mathrm{c}^{\,\mathrm{\max}}$ \\
  &  [$M_\odot$]  &  [$M_\odot$] &  [$M_\odot$] & [$10^{\,9}~{\mathrm{K}}$] & [$\mathrm{g~cm^{-3}}$] & &  \\    
\hline
\hline
150~$M_\odot$ & 94 & 72 & 64 & 3.45 & 6.25 & $1.0\times10^{\,-4}$  & $2.5\times 10^{\,-4}$ \\
70~$M_\odot$ He & 70 & 70 & ~60 & 3.57 & 6.30 & $1.9\times10^{\,-7}$ &  $2.8\times 10^{\,-4}$ \\
\hline
250~$M_\odot$ & 169 & 121 & 110 & 5.12 & 6.69 &$1.0\times10^{\,-4}$   & $1.6\times 10^{\,-3}$ \\
115~$M_\odot$ He & 115 & 115 & ~90 & 5.14 & 6.67 & $1.9\times10^{\,-7}$ &  $7.3\times 10^{\,-4}$ \\
\hline
\end{tabular}
\end{center}
\end{table*}  

\section{Results}
\label{sect:result1}

\subsection{Explosion}
\label{subsect:PIresults}

Through the hydrodynamics terms included in the BEC code (see the Appendix~\ref{appendix:append}) we could follow
the dynamical phase of the evolution of our PISN models.  Usually hydrodynamic stellar evolution codes are not able to 
describe dynamical processes in stars because of the implicit nature of the
adopted numerical solvers (causing strong numerical damping) and the large time steps which are required to follow the 
evolution time scale \citep{1970A&A.....9..216A,1982sscr.conf...79W}.  
Pulsations and shock waves however can be resolved if the time step becomes comparable to the dynamical characteristic time (possible
during late stages of stellar evolution), and if the growth rate of a hydrodynamical phenomenon is sufficiently large
\citep{1986A&A...167..274E,1997A&A...327..224H,2010ApJ...717L..62Y}.

\begin{figure}
\centering
\includegraphics[width=\columnwidth]{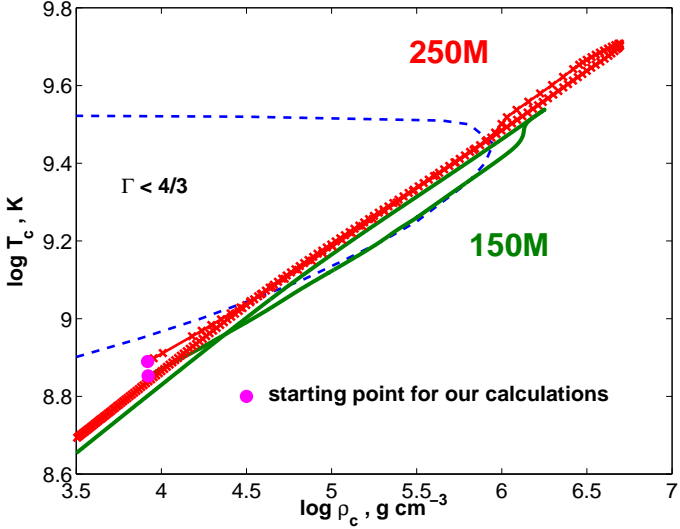}
\caption[Evolutionary tracks of our 150~$M_\odot${} and 250~$M_\odot${} models in central density~--~temperature diagram.]
{Evolutionary tracks of our 150~$M_\odot${} (labeled `150M', solid line) and 250~$M_\odot${} 
(labeled `250M', line with times signs) models in central density~--~temperature
diagram.  The area enclosed with the dashed line indicates the pair instability regime where $\Gamma~<~4/3$. 
The filled circles mark the starting points for each model sequence.}
\label{figure:rhoTc}
\end{figure}

\begin{figure}[h]
\centering
\includegraphics[width=\columnwidth]{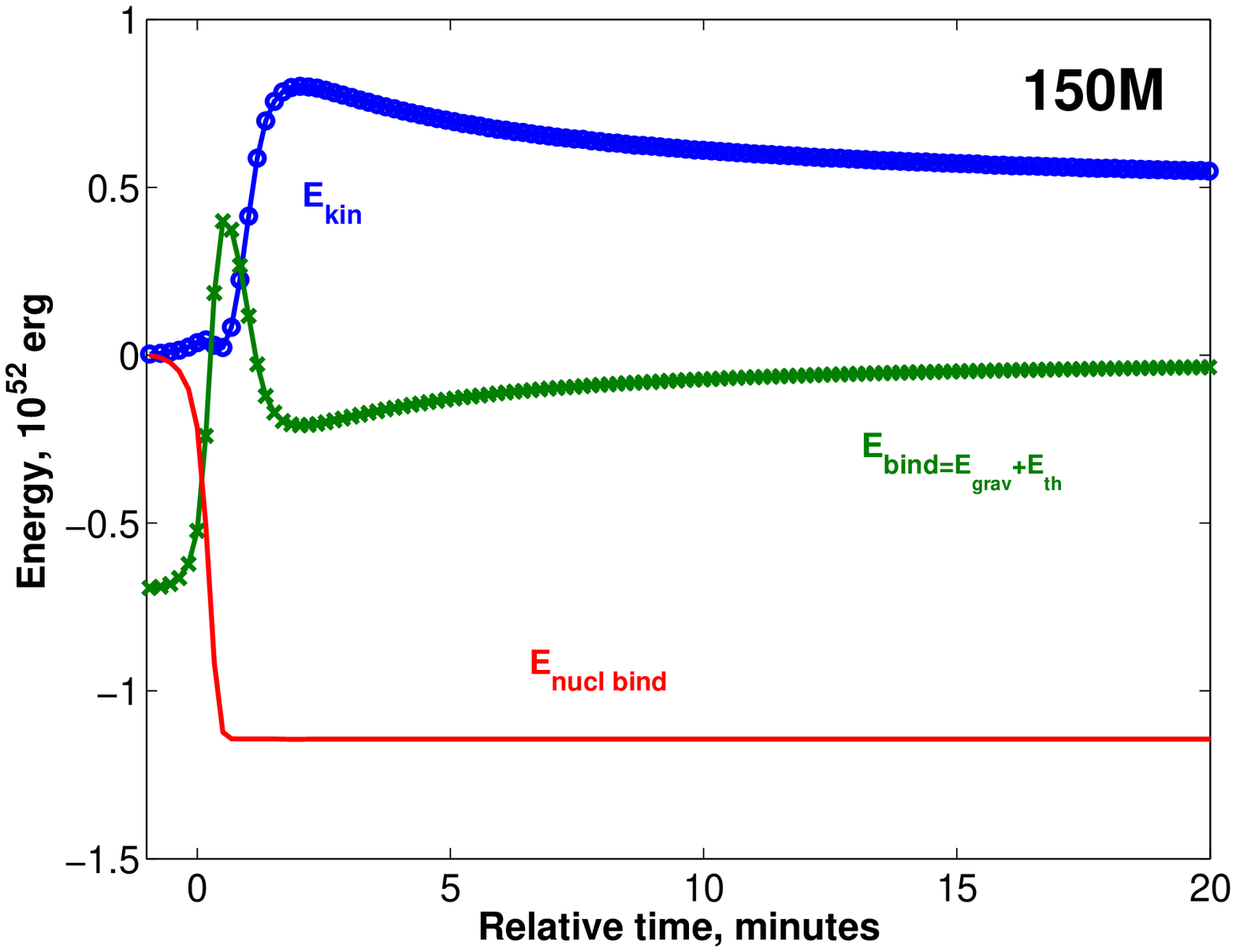}
\includegraphics[width=\columnwidth]{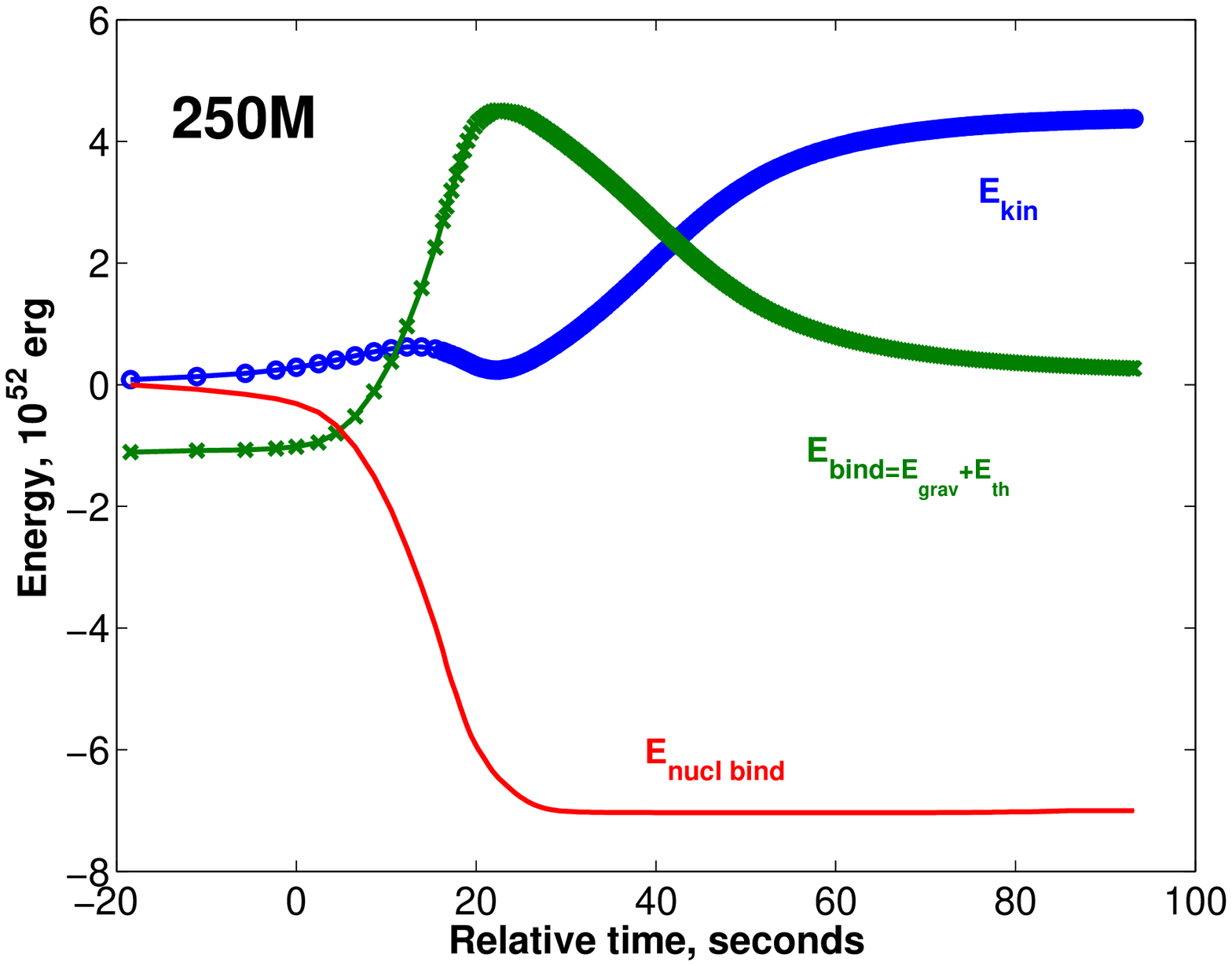}
\caption[The energetics of the PISN explosions for our 150~$M_\odot${} and 250~$M_\odot${} models.]
{Evolution of the energetics for our 150~$M_\odot${} and 250~$M_\odot${} models.  The kinetic energy (blue, circles),  
the binding energy (green, times) that is the sum of gravitational and thermal energies, and the nuclear
binding energy (red) of the stars are shown.  
Here, the nuclear binding energy is defined by the difference between the total nuclear binding energy of all nuclei at the
end of carbon burning and that of a given evolutionary epoch (see description in the text).  
The zero point in time is defined as the time of the beginning of pair instability explosion.}
\label{figure:energetics}
\end{figure}

We find that both models explode as a result of explosive nuclear burning
during the pair instability phase, which confirms the prediction by
\citet{2007A&A...475L..19L}.  

\begin{figure}[h]
\centering
\includegraphics[width=\columnwidth]{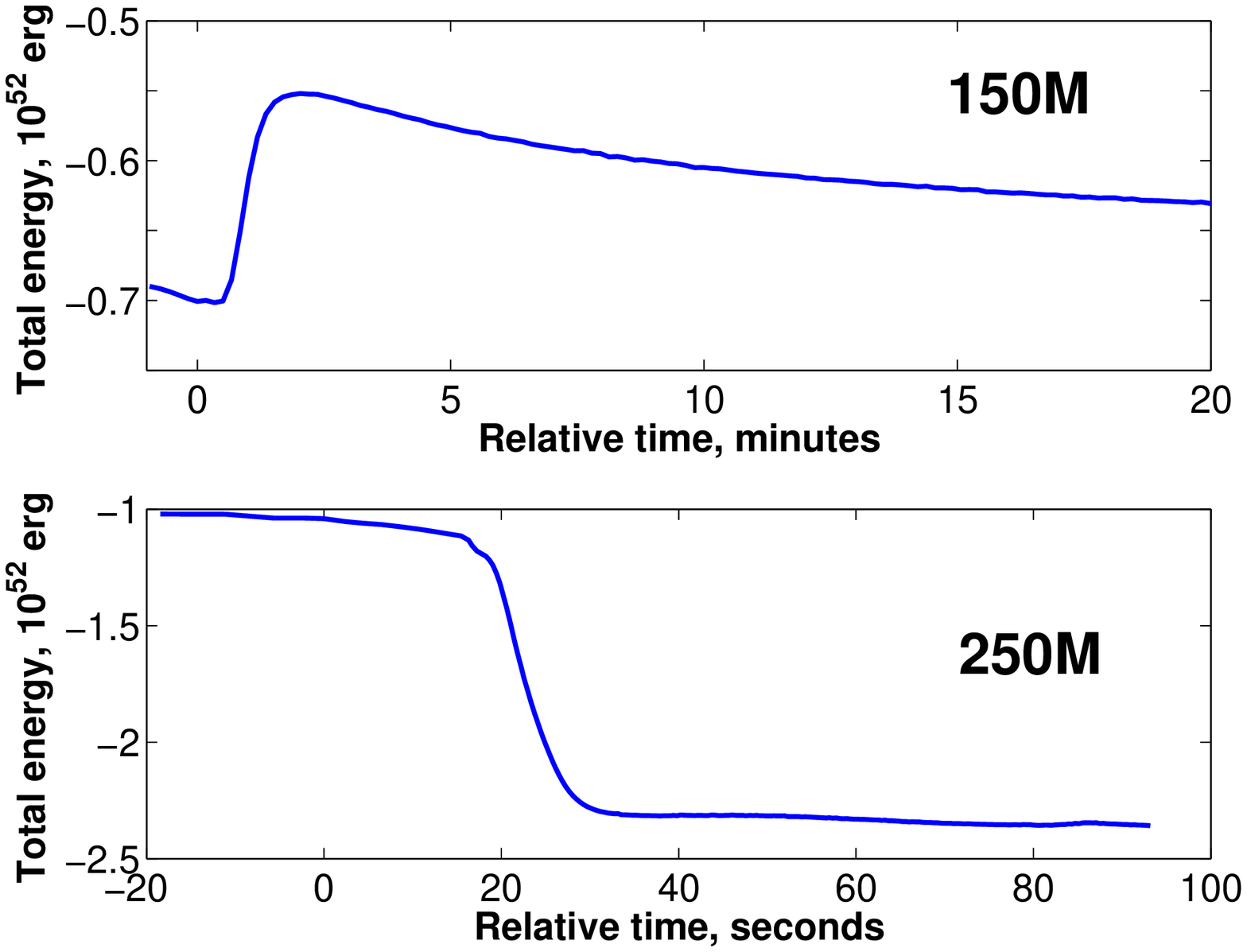}
\caption[Total energy evolution for our 1500~$M_\odot${} and 250~$M_\odot${} models.]
{Total energy evolution for our 1500~$M_\odot${} (upper) and 250~$M_\odot${} (bottom) models.}
\label{figure:etot}
\end{figure}

In Figure~\ref{figure:rhoTc}, the evolutionary
tracks of the  central density and temperature are shown.  Both quantities
increase rapidly during the dynamical contraction of the core induced by the
pair instability.  The maximum temperature and density achieved during this
phase are $T_{\mathrm{c}}  =  3.45 \times 10^{\,9}~\mathrm{K}$ and
$\rho_{\mathrm{c}} = 1.8 \times 10^{\,6}~\mathrm{g~cm^{-3}}$ for the 150~\Msun{} star,
and $T_{\mathrm{c}}  = 5.1 \times 10^{\,9}~\mathrm{K}$ and  $\rho_{\mathrm{c}} =
4.9 \times 10^{\,6}~\mathrm{g~cm^{-3}}$ for the 250~\Msun{} star, respectively.  As
shown in Table~\ref{table:modeldata}, these values are comparable to those
found in Population~III star models by \citet{2002ApJ...567..532H}
for similar helium core masses.  Beyond this point, the
contraction is reversed, and the star explodes.  

In Figure~\ref{figure:energetics}, the fact that our models explode by the
nuclear energy release is illustrated.  Initially our models have a negative
binding energy 
($E_\mathrm{bind} = E_\mathrm{grav} + E_\mathrm{thermal}$).   The pair creation
triggers the collapse which is visible as a minor increase of the kinetic energy
around $t = 0$ for Model~150M and around $t = 0 - 10$~s for Model~250M. The
consequent oxygen and silicon burning occur with a sharp decrease of the nuclear
binding energy.  

The nuclear binding energy is defined the following way:
\begin{equation}
E_\mathrm{bind} = \sum_{k} \sum_{i} X_i \, e_i \, {\Delta M_k\over{m_i}}\, ,
\label{equation:ebindnucl}
\end{equation}
where $X_i$ represents the mass fractions of the isotopes, $e_i$ and $m_i$ are the nuclear binding energy and nuclear mass of the isotope
$i$, and $\Delta M$ is the mass of the mass shell of the numerical stellar model.  The sum is made over all mass shells and over all isotopes.  
Figure~\ref{figure:energetics} shows the difference of nuclear binding energy at a given time to that at the
end of carbon burning, i.e. the first time point in the figures.  The drop in the nuclear binding energy shows the amount of
energy released by nuclear burning during the pair instability explosion.  

The released nuclear energy from oxygen and silicon burning is
converted into thermal and kinetic energy, resulting in a positive binding
energy.  A positive binding energy means that the system becomes unbound.
Eventually, most of the released nuclear energy is converted into the kinetic
energy,  which is strong evidence for the explosion of the star.  The final
kinetic energy is 8~foe\footnote{1 foe = $10^{\,51}$erg (from `fifty-one-erg')} for Model~150M and 44~foe for Model~250M.  This
corresponds to an 
asymptotic velocity of the ejecta at the infinity of $2.9 \times 10^{\,3}$
km~s$^{\,-1}$ for Model~150M and $5.1 \times 10^{\,3}$ km~s$^{\,-1}$ for
Model~250M.  

%

The BEC code has no provision for treating shocks.  However, due to the strong density contrast at the base of the hydrogen-rich
envelope, a shock wave develops at this point due to the explosion of the carbon-oxygen core.  
Its Mach number is about 2 for both our models.  
As a consequence, energy is
not perfectly conserved in our models at the time when the shock enters the stellar envelope (see Figure\ref{figure:etot}).  
Still the total energy is conserved to better than 5\% in our 150~\Msun model, and to better than 20\% in our 250~\Msun model.  
We note that the velocities scale with $\sqrt{E_{\mathrm{kin}}}$, and that we expect our velocities to be precise to 3\% and
10\%, respectively.  The comparison of Figures~\ref{figure:energetics} and \ref{figure:etot} shows that energy conservation during the
nuclear burning phase is very good, such that our nucleosynthesis results are not affected by this issue.

\subsection{Nucleosynthesis}
\label{subsect:nucresults}

\begin{figure}
\centering
\includegraphics[width=\columnwidth]{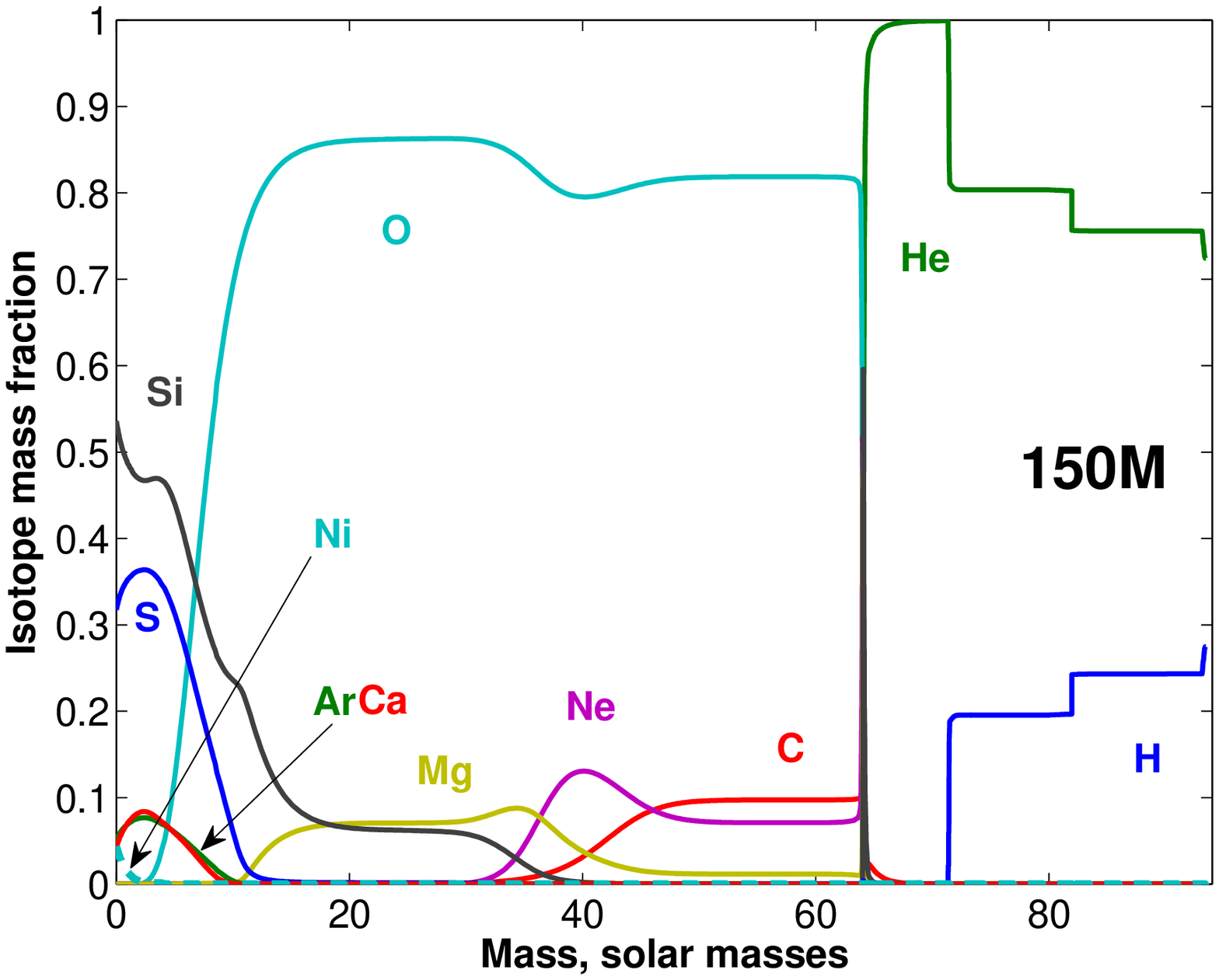}
\includegraphics[width=\columnwidth]{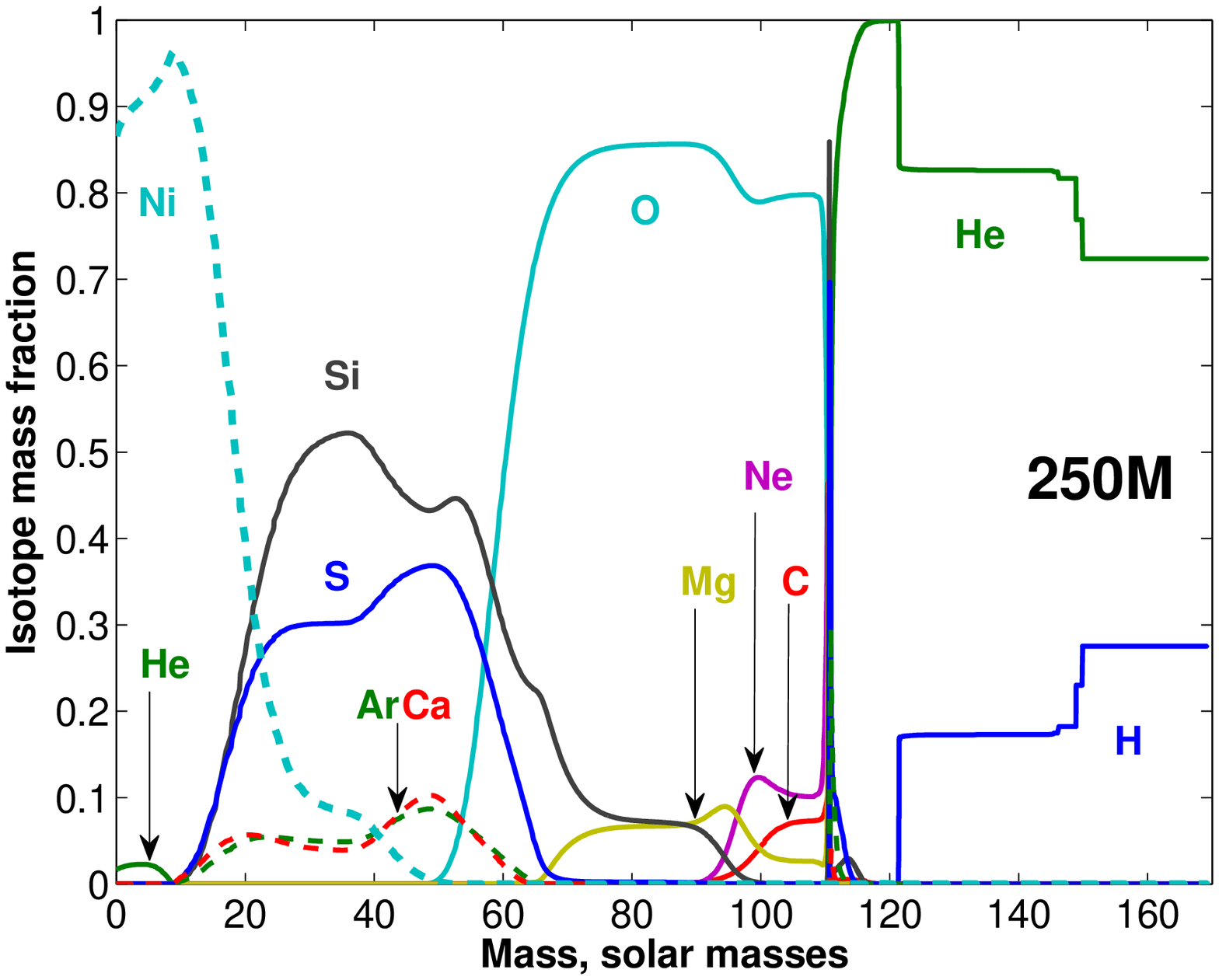}
\caption[The final chemical structure of our models.]
{The final chemical structure of our models.}
\label{figure:iso}
\end{figure}

\begin{table*}
\caption[Total nucleosynthetic yields for selected isotopes in solar masses for our 150~$M_\odot${} and 250~$M_\odot${} models 
in comparison with 70~$M_\odot${} and 115~$M_\odot${} zero metallicity helium star yields.]
{Total nucleosynthetic yields for selected isotopes in solar masses for our 150~$M_\odot${} and 250~$M_\odot${} models
(including matter lost by stellar wind and decay products)
in comparison with 70~$M_\odot${} and 115~$M_\odot${} zero metallicity helium star yields \citep{2002ApJ...567..532H}, 
respectively.  Yields for hydrogen and helium neglecting the wind matter are given after the slash sign.}
\begin{center}
\begin{tabular}{rcccccccccccccc}
\hline
\, & $Z$ & $^1$H & $^4$He & $^{12}$C & $^{16}$O & $^{20}$Ne & $^{24}$Mg & $^{28}$Si & $^{32}$S & $^{36}$Ar & $^{40}$Ca & $^{46}$Ti & $^{50}$Cr &$^{56}$Fe \\
\hline
150~$M_\odot$ & 0.001 & 36.6/4.9 & 49.2/24.4 & 2.2 & 46.9 & 2.6 & 2.3 & 6.2  & 2.8  & 0.5 & 0.5 & $10^{-4}$ & $10^{-4}$ & 0.04 \\
He 70~$M_\odot$ & 0 & - & 1.5 & 4.5 & 45.8 & 4.0 & 3.0 & 8.0 & 2.4 & 0.3 & 0.2 & $10^{-4}$ & $10^{-4}$ & 0.01 \\
\hline
250~$M_\odot$ & 0.001 & 57.1/10.3 & 81.5/47.5 & 0.9 & 42. & 1.8 & 2.5 & 23.1 & 14.3 & 2.9 & 2.8 & $10^{-4}$ & $10^{-4}$ & 19.4 \\
He 115~$M_\odot$ & 0 & - & 1.8 & 3.7 & 40. & 3.8 & 4.5 & 25.7 & 11.8 & 1.9 & 1.6 & $10^{-5}$ & $10^{-4}$ & 19.0 \\
\hline
\end{tabular}
\label{table:yields_selected}
\end{center}
\end{table*}

\begin{table*}
\caption[Total nucleosynthetic yields in solar masses and production factors for our 150~$M_\odot${} and 250~$M_\odot${} models.]
{Total nucleosynthetic yields in solar masses and production factors for our 150~$M_\odot${} and 250~$M_\odot${} models.
Yields include matter lost by the stellar wind and decay products.}
\begin{center}
\begin{tabular}{r|rr|rr||r|rr|rr}
\hline
\, & \multicolumn{2}{c|}{Yields [\Msun]} & \multicolumn{2}{c||}{Prod. factor} & 
\, & \multicolumn{2}{c|}{Yields [\Msun]} & \multicolumn{2}{c}{Prod. factor} \\
\hline
\, & 150~$M_\odot$ & 250~$M_\odot$ & 150~$M_\odot$ & 250~$M_\odot$ &
\, & 150~$M_\odot$ & 250~$M_\odot$ & 150~$M_\odot$ & 250~$M_\odot$ \\
\hline
$^1$H     & 36.56     & 57.10     & 0.34      & 0.32      &
$^{41}$K  & 1.63(-4)  & 3.15(-4)  & 3.63      & 4.22      \\
$^2$H     & 7.10(-11) & 1.63(-10) & 1.71(-8)  & 2.36(-8)  &
$^{40}$Ca & 0.49      & 2.80      & 45.99     & 156.82    \\
$^3$He    & 5.01(-5)  & 5.47(-5)  & 9.80(-3)  & 6.42(-3)  &
$^{42}$Ca & 1.87(-4)  & 2.09(-4)  & 2.50      & 1.67      \\ 
$^4$He    & 49.16     & 81.53     & 1.20      & 1.19      &
$^{43}$Ca & 6.48(-8)  & 2.07(-5)  & 4.30(-3)  & 0.78      \\
$^6$Li    & 4.93(-12) & 2.10(-11) & 4.45(-5)  & 1.14(-4)  &
$^{44}$Ca & 1.14(-4)  & 9.65(-4)  & 0.45      & 2.28      \\
$^7$Li    & 7.17(-10) & 4.57(-8)  & 4.55(-4)  & 1.74(-2)  &
$^{46}$Ca & 0         & 0         & 0         & 0         \\
$^9$Be    & 8.77(-12) & 1.88(-11) & 3.01(-4)  & 3.88(-4)  &
$^{48}$Ca & 8.71(-20) & 3.92(-17) & 3.50(-15) & 9.45(-13) \\
$^{10}$B  & 2.14(-10) & 4.25(-10) & 1.43(-3)  & 1.70(-3)  &
$^{45}$Sc & 4.32(-6)  & 1.18(-5)  & 0.64      & 1.05      \\
$^{11}$B  & 7.17(-10) & 4.58(-9)  & 1.07(-3)  & 4.10(-3)  &
$^{46}$Ti & 9.23(-5)  & 1.14(-4)  & 2.29      & 1.69      \\
$^{12}$C  & 2.22      & 0.90      & 6.04      & 1.46      &
$^{47}$Ti & 3.87(-7)  & 2.42(-6)  & 1.04(-2)  & 3.92(-2)  \\
$^{13}$C  & 3.58(-4)  & 5.60(-4)  & 8.00(-2)  & 7.52(-2)  &
$^{48}$Ti & 2.02(-4)  & 1.69(-2)  & 0.54      & 26.96     \\
$^{14}$N  & 4.10(-2)  & 7.27(-2)  & 0.34      & 0.37      &
$^{49}$Ti & 1.91(-5)  & 4.61(-4)  & 0.68      & 9.81      \\
$^{15}$N  & 1.17(-4)  & 3.37(-5)  & 0.25      & 4.31(-2)  &
$^{50}$Ti & 4.23(-12) & 3.35(-12) & 1.53(-7)  & 7.29(-8)  \\
$^{16}$O  & 46.86     & 41.96     & 47.33     & 25.43     &
$^{50}$V  & 1.63(-10) & 1.65(-10) & 1.04(-3)  & 6.28(-4)  \\
$^{17}$O  & 1.10(-4)  & 1.54(-4)  & 0.28      & 0.24      &
$^{51}$V  & 1.93(-5)  & 6.42(-4)  & 0.30      & 5.98      \\
$^{18}$O  & 1.24(-4)  & 4.94(-5)  & 5.57      & 1.33(-2)  &
$^{50}$Cr & 2.43(-4)  & 1.02(-3)  & 1.98      & 4.99      \\
$^{19}$F  & 3.76(-7)  & 4.03(-7)  & 5.36      & 3.45(-3)  &
$^{52}$Cr & 1.62(-3)  & 0.38      & 0.66      & 91.77     \\
$^{20}$Ne & 2.58      & 1.85      & 14.69     & 6.32      &
$^{53}$Cr & 1.79(-4)  & 1.40(-2)  & 0.63      & 29.60     \\
$^{21}$Ne & 4.17(-4)  & 8.79(-5)  & 0.95      & 0.12      &
$^{54}$Cr & 1.09(-8)  & 2.02(-8)  & 1.51(-4)  & 1.69(-4)  \\
$^{22}$Ne & 1.19(-3)  & 1.17(-3)  & 8.42(-2)  & 4.94(-2)  &
$^{55}$Mn & 1.00(-3)  & 5.73(-2)  & 0.45      & 15.59     \\
$^{23}$Na & 1.39(-2)  & 8.67(-3)  & 2.39      & 0.90      &
$^{54}$Fe & 2.12(-2)  & 0.21      & 1.82      & 11.07     \\
$^{24}$Mg & 2.30      & 2.52      & 27.12     & 17.83     &
$^{56}$Fe & 4.62(-2)  & 19.33     & 0.24      & 61.36     \\
$^{25}$Mg & 1.34(-2)  & 5.37(-3)  & 1.20      & 0.29      &
$^{57}$Fe & 4.41(-4)  & 0.21      & 9.93(-2)  & 28.38     \\
$^{26}$Mg & 2.68(-2)  & 1.11(-2)  & 2.09      & 0.52      &
$^{58}$Fe & 4.09(-6)  & 5.44(-5)  & 6.81(-3)  & 5.44(-2)  \\
$^{27}$Al & 1.14(-2)  & 4.20(-2)  & 1.15      & 2.53      &
$^{59}$Co & 3.99(-5)  & 5.11(-3)  & 6.63(-2)  & 5.10      \\
$^{28}$Si & 6.17      & 23.08     & 54.52     & 122.29    &
$^{58}$Ni & 1.83(-3)  & 0.37      & 0.22      & 26.73     \\
$^{29}$Si & 4.62(-2)  & 3.24(-2)  & 7.76      & 3.26      &
$^{60}$Ni & 1.24(-4)  & 7.35(-2)  & 3.75(-2)  & 13.36     \\
$^{30}$Si & 2.52(-2)  & 1.35(-2)  & 6.19      & 2.00      &
$^{61}$Ni & 2.57(-7)  & 3.75(-3)  & 1.76(-3)  & 15.44     \\
$^{31}$P  & 2.96(-3)  & 8.16(-3)  & 2.60      & 4.30      &
$^{62}$Ni & 9.40(-7)  & 2.59(-2)  & 1.99(-3)  & 32.86     \\
$^{32}$S  & 2.82      & 14.32     & 47.47     & 144.69    &
$^{64}$Ni & 7.44(-12) & 4.46(-9)  & 6.00(-8)  & 2.16(-5)  \\
$^{33}$S  & 3.83(-3)  & 8.14(-3)  & 7.93      & 10.11     &
$^{63}$Cu & 3.95(-9)  & 2.40(-5)  & 3.92(-5)  & 0.14      \\
$^{34}$S  & 8.17(-3)  & 1.12(-2)  & 2.93      & 2.41      &
$^{65}$Cu & 1.76(-11) & 9.92(-6)  & 3.80(-7)  & 0.13      \\
$^{36}$S  & 2.36(-8)  & 5.23(-8)  & 1.96(-3)  & 2.62(-3)  &
$^{64}$Zn & 1.98(-9)  & 1.20(-4)  & 1.18(-5)  & 0.43      \\
$^{35}$Cl & 7.24(-4)  & 1.09(-2)  & 1.19      & 10.70     &
$^{66}$Zn & 6.34(-11) & 2.24(-4)  & 6.40(-7)  & 1.36      \\
$^{37}$Cl & 6.68(-4)  & 1.37(-3)  & 3.25      & 4.01      &
$^{67}$Zn & 5.63(-15) & 1.56(-7)  & 3.81(-10) & 6.34(-3)  \\
$^{36}$Ar & 0.51      & 2.93      & 37.28     & 128.15    &
$^{68}$Zn & 8.90(-15) & 4.64(-8)  & 1.30(-10) & 4.06(-4)  \\
$^{38}$Ar & 6.75(-3)  & 7.92(-3)  & 2.57      & 1.81      &
$^{70}$Zn & 0         & 0         & 0         & 0         \\
$^{40}$Ar & 1.37(-10) & 3.11(-9)  & 3.26(-5)  & 4.42(-4)  &
$^{69}$Ga & 4.22(-16) & 4.47(-13) & 6.45(-11) & 4.10(-8)  \\
$^{39}$K  & 1.23(-3)  & 1.16(-2)  & 2.09      & 11.82     &
$^{71}$Ga & 0         & 0         & 0         & 0         \\
$^{40}$K  & 3.71(-8)  & 1.68(-6)  & 3.95(-2)  & 1.07      &
$^{70}$Ge & 6.93(-16) & 9.16(-15) & 8.82(-11) & 7.00(-10) \\
\hline
\end{tabular}
\label{table:yields}
\end{center}
\end{table*}

\begin{figure}
\centering
\includegraphics[width=\columnwidth]{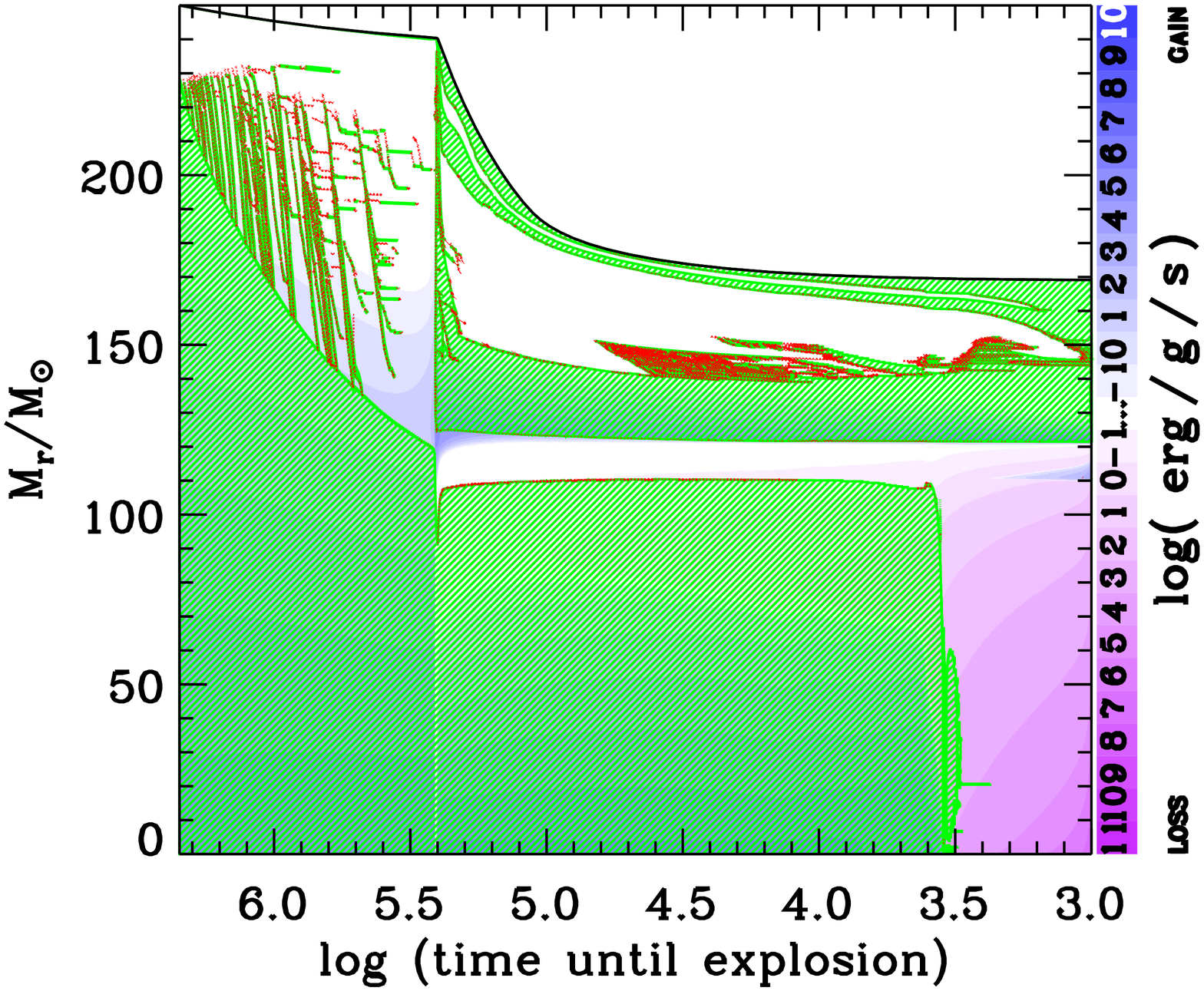}
\caption[Kippenhahn diagram for 250~$M_\odot${} PISN model.]{Kippenhahn diagram for the 250~$M_\odot${} PISN model.  
Convective and semi-convective layers are marked by green hatched lines
and red dots, respectively.  The net amount of local energy loss and production is indicated
by colar shading.  The surface of the star is marked by the black solid line. 
}\label{figure:kipp}
\end{figure}

Figure~\ref{figure:iso} shows the final chemical structure of our
models.  The total amounts of produced $^{56}$Ni are 0.04~$M_\odot${} and
19.3~$M_\odot${} for the 150~$M_\odot${} and 250~$M_\odot${} star, respectively.  As shown
in Table~\ref{table:yields_selected}, the overall nucleosynthetic results are
in good agreement with that of the 70~\Msun{} and 115~\Msun{} Population~III helium star models
by \citet{2002ApJ...567..532H} of which the masses are comparable to the He
core masses of our models.

We summarize the chemical yields and the production factors of each isotope from
our explosion models in Table~\ref{table:yields}.  
Here the production factor of a given isotope is defined as  
\begin{equation}\label{equation:eq1}
p_{\,\mathrm{iso}} = \frac{m_\mathrm{iso}}{{X^{\,\odot}_{\mathrm{iso}} \times
M_{\,\mathrm{tot}}}}~,   
\end{equation}
where $m_\mathrm{iso}$ is the total yield of a given isotope in solar masses, 
$X^{\,\odot}_{\mathrm{iso}}$ is the mass fraction of the isotope according to the solar metallicity pattern, and
$M_{\,\mathrm{tot}}$ is the initial mass of the star (150~\Msun{} or 250~\Msun{} in
the present study).  The effect of radioactive decays is fully considered in the
final set of the isotope yields that consists only of stable nuclei.

\begin{figure}
\centering   
\includegraphics[width=0.43\textwidth]{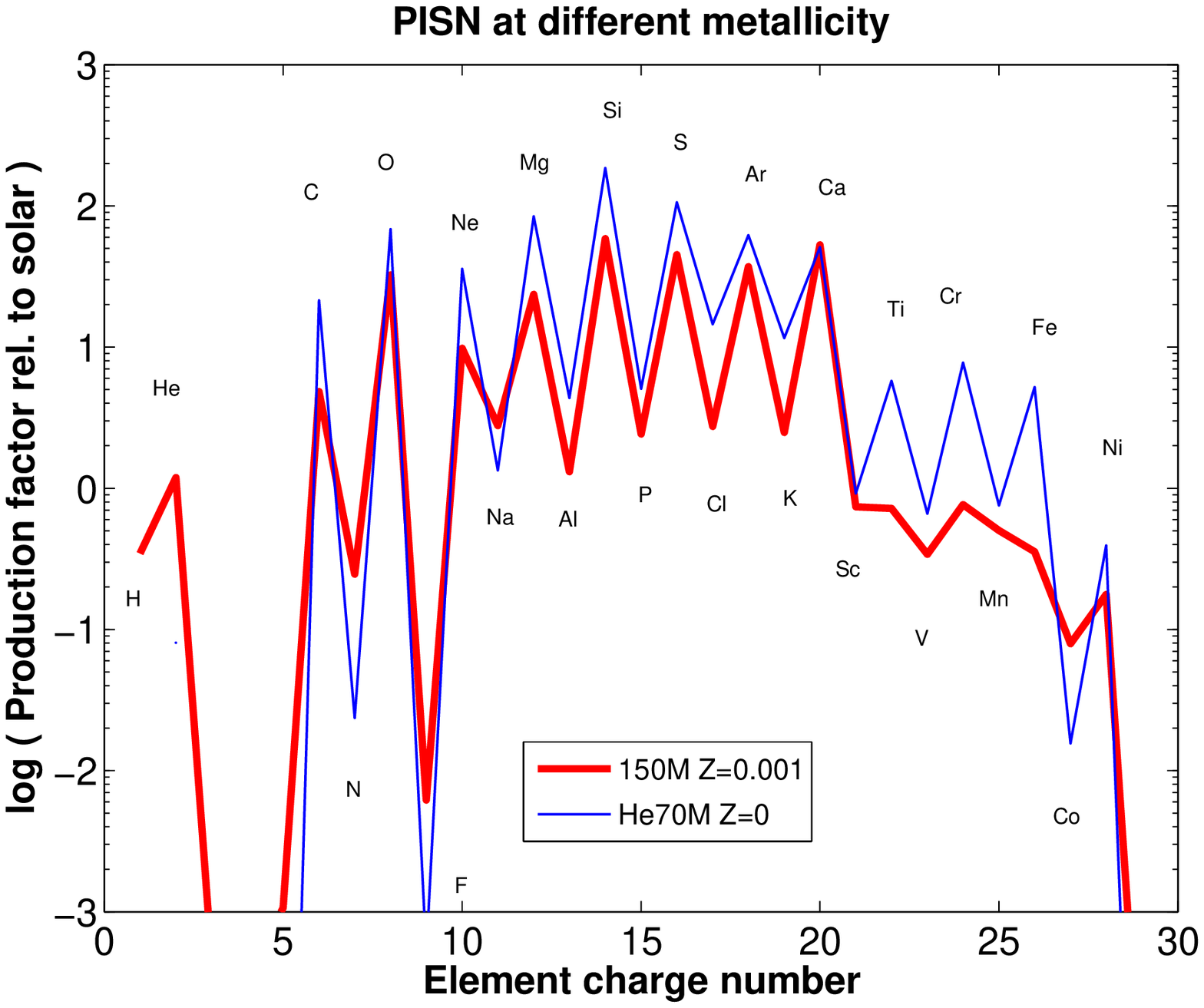}
\includegraphics[width=0.43\textwidth]{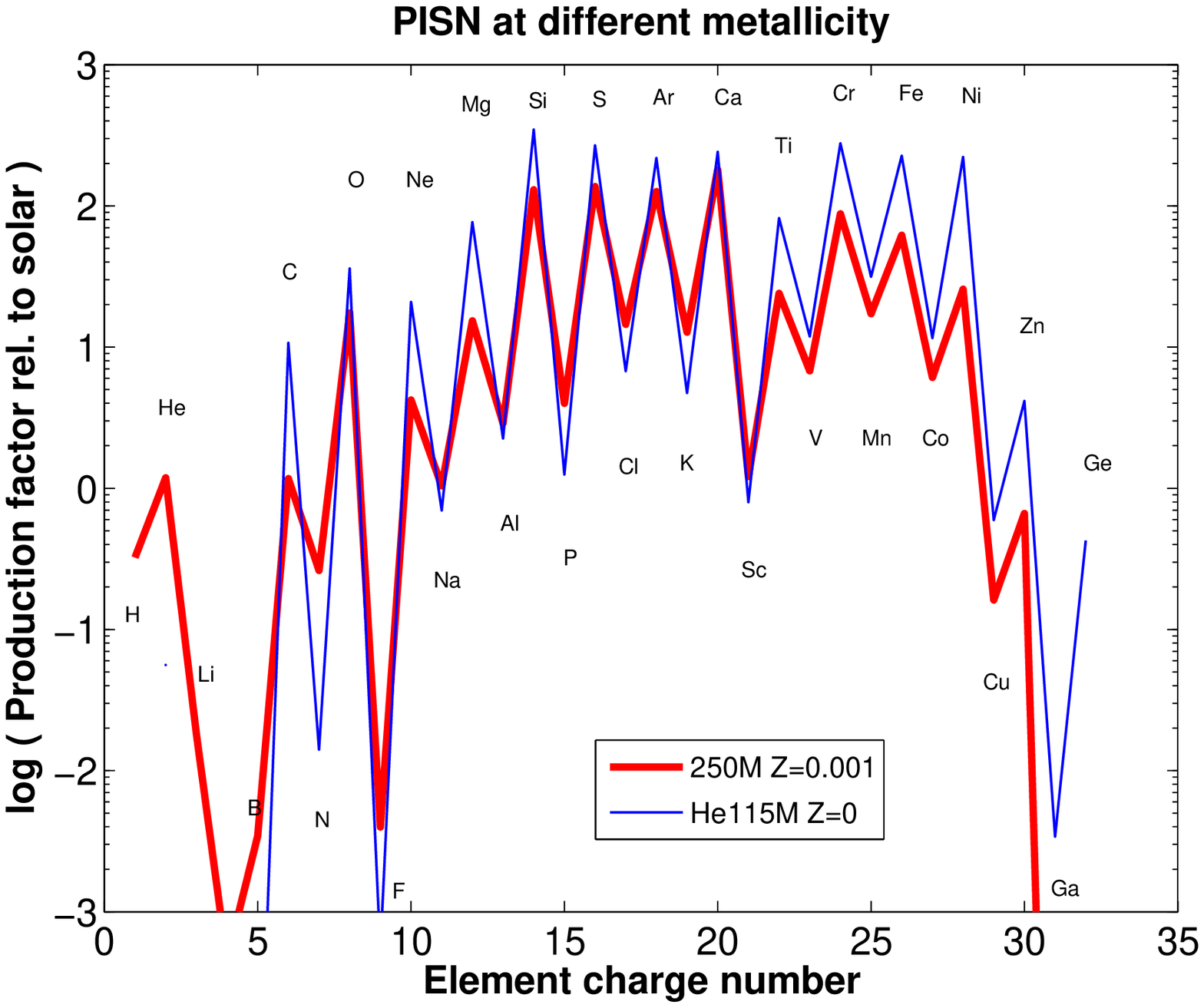}
\caption[Production factors of major elements from our 150~\Msun and 250~\Msun{} PISN  
models with those of comparable Population~III helium star models.]
{Production factors of major elements from our 150~\Msun (upper panel) and 250~\Msun{}(lower panel) PISN
models (red thick lines) compared with those of comparable 70~\Msun{} (upper panel) and 115~\Msun{} (lower panel) 
Population~III helium star model by \citet{2002ApJ...567..532H} (blue thin lines).}
\label{figure:prod1a}
\end{figure}

\begin{figure}
\centering   
\includegraphics[width=0.43\textwidth]{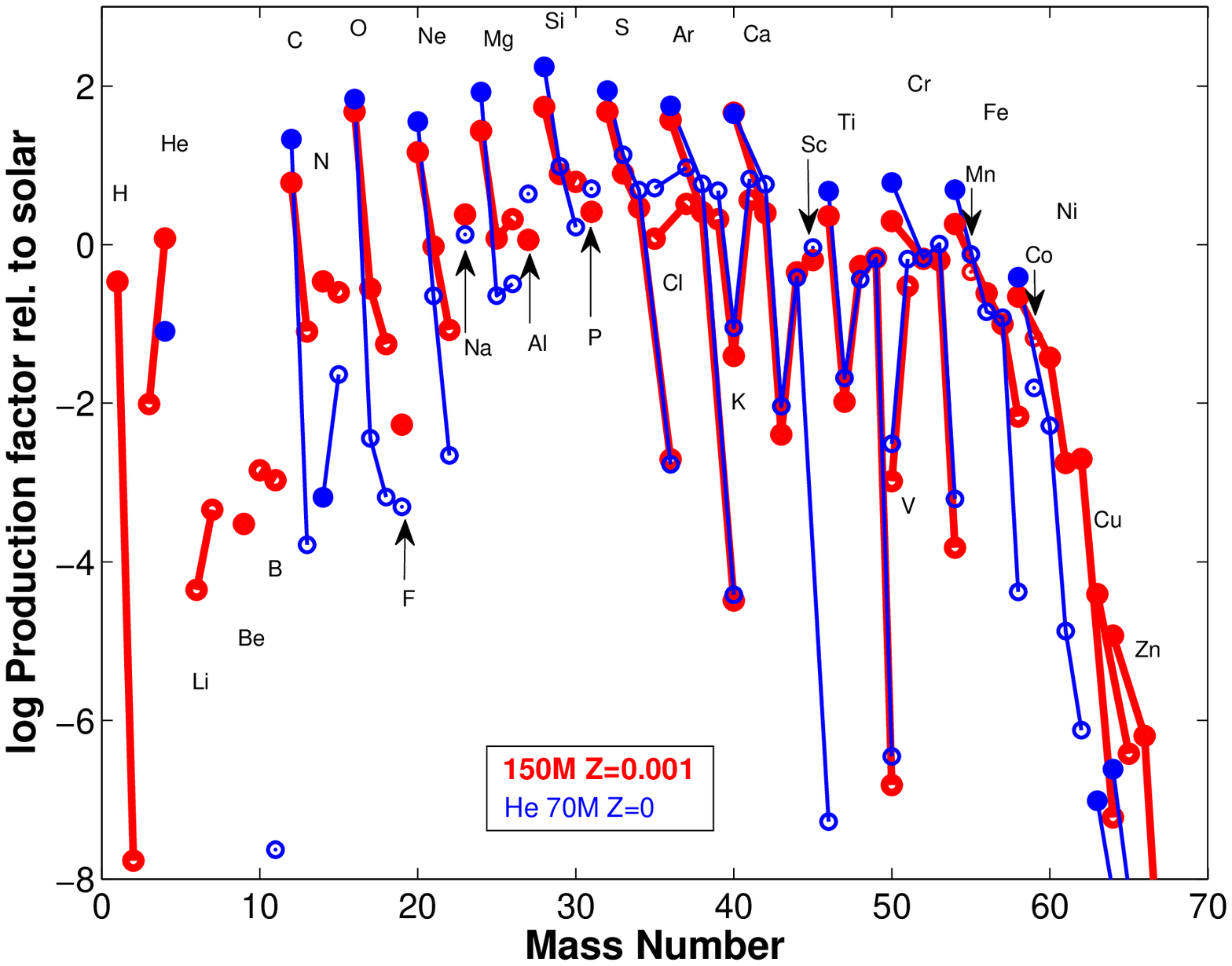}
\includegraphics[width=0.43\textwidth]{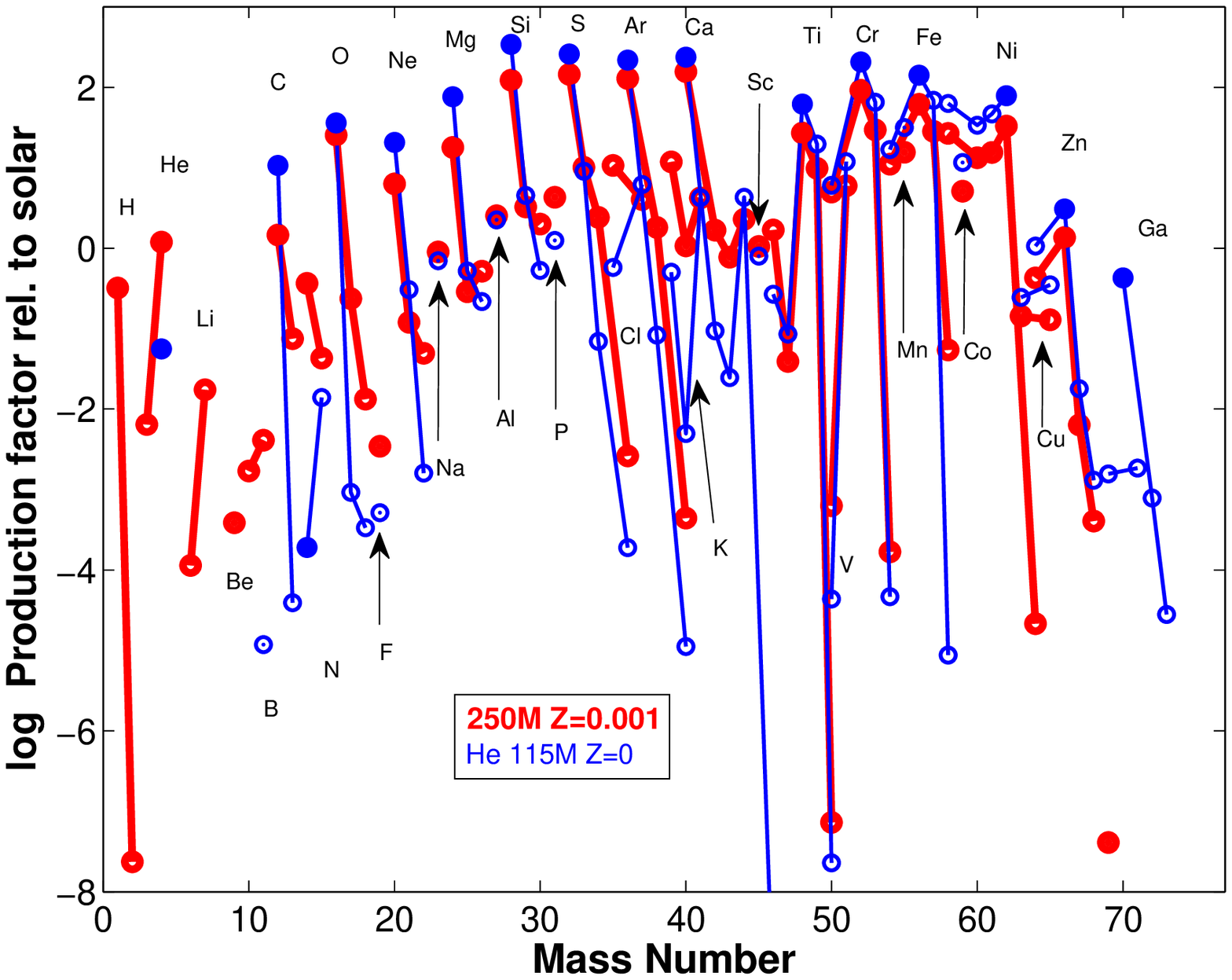}
\caption[Isotopic production factors for the indicated nuclei for our 150~\Msun and 250~\Msun{} PISN models for those of comparable
Population~III helium star models.]
{Isotopic production factors for the indicated nuclei.  
The isotopes of a given element are connected by solid lines.  The filled circles indicates the most abundant isotope
for each element, while the open circles denote the other isotopes.  
Our 150~\Msun (upper panel) and 250~\Msun{}(lower panel) PISN models (red thick) are compared 
with those of comparable 70~\Msun{} (upper panel) and 115~\Msun{} (lower panel)
Population~III helium star model by \citet{2002ApJ...567..532H} (blue thin).}
\label{figure:prod1b}
\end{figure}

We use the solar abundances taken from \citet{1995ApJS..101..181W} which were adopted from \citet{1989GeCoA..53..197A} to be consistent
in our comparison to other PISN and CCSN nucleosynthetic yields.  A more recent
study of solar abundances \citep{2009ARA&A..47..481A} shows somewhat different solar abundances.  The overall fraction of heavy elements
differs by a factor of 0.7 mostly due to a reduced oxygen abundance. $Z=0.0134$ in \citet{2009ARA&A..47..481A} and $Z=0.0201$ in \citet{1989GeCoA..53..197A}.  However, the solar
abundances serve as a denominator for our qualitative comparison.  Generally, the relative scatter of the elemental/isotopic 
production factor (i.e. odd-even effect) remains the same.  We plot all production factors in
Figures~\ref{figure:prod1a}, \ref{figure:prod1b}, and \ref{figure:prod2b} in logarithmic scale.  Therefore, using the lower metal
fraction will shift all data (except hydrogen and helium) by $| \log 0.7 | \simeq 0.15$~dex.  

The pair instability explosion in our models is mostly driven by oxygen burning because oxygen is the
most abundant element at core carbon exhaustion.  Our models 150M and 250M
contain 64~\Msun{} and 110~\Msun{} oxygen cores, correspondingly, and 
a large fraction of the oxygen core remains
unburnt: more than 40~\Msun{} of oxygen enrich the circumstellar medium, making
oxygen the third most abundant element after hydrogen and
helium.  Note that only about 3~\Msun{} of oxygen are produced in an ordinary
core-collapse SN \citep{1995ApJS..101..181W} and even less (about 0.1~\Msun{})
is left after a SN~Ia \citep{1999ApJS..125..439I,2004A&A...425.1029T}.

The highest yields in Model~150M are those of intermediate
even-charged isotopes between oxygen and sulphur (2.6~\Msun{} of $^{20}$Ne,
2.3~\Msun{} of $^{24}$Mg, 6.2~\Msun{} of $^{28}$Si, 2.8~\Msun{} of $^{32}$S)
because only a small fraction of silicon is burnt in this PISN (see
Table~\ref{table:yields}).  The yields of iron-group isotopes are fairly low
compared to intermediate mass isotopes.  At the same time the ejecta of Model~250M
contains large amounts of intermediate mass isotopes similar to the Model~150M
(1.85~\Msun{} of $^{20}$Ne, 2.5~\Msun{} of $^{24}$Mg, 23.1~\Msun{} of
$^{28}$Si, 14.3~\Msun{} of $^{32}$S, 2.9~\Msun{} of $^{36}$Ar, 2.8~\Msun{} of
$^{40}$Ca), significant amounts of iron-group isotopes (0.4~\Msun{} of
$^{52}$Cr, 0.2~\Msun{} of $^{54}$Fe, 0.2~\Msun{} of $^{57}$Fe, 0.4~\Msun{} of
$^{58}$Ni) and a very large iron-56 yield (19.3~\Msun). There is a gap around
the titanium isotopes since these are the bottle-neck isotopes between
QSE-groups
\citep{1996ApJ...460..869H}.

Large amounts of silicon (6~\Msun{} and 23~\Msun{}, respectively)  are left
after incomplete silicon burning in both models, which are 10~--~100~times
higher than silicon yields resulting from core-collapse SNe and SNe~Ia.  The 
average yield of silicon in core-collapse SNe and SNe~Ia is 0.4~\Msun{} and
0.6~\Msun{}, respectively.  A large amount of radioactive nickel produced in
our higher mass Model~250M (19.3~\Msun) will result in a very bright and broad
supernova light curve~\citep{2005ApJ...633.1031S,2011ApJ...734..102K,2014arXiv1403.5212K}.  
This amount is much larger than the
average iron yield of 0.04~--~0.2~\Msun{} in core-collapse SNe and
0.5~\Msun{} in SNe~Ia \citep{1994A&A...282..731P,2009ARA&A..47...63S}.  

We emphasize here that both our PISN models do not produce pronounced amounts of
isotopes beyond the iron-group: the mass fractions of copper, zinc gallium and
germanium isotopes are well below $10^{\,-4}$.  The lack of {\emph r}- and {\emph
s}-isotopes is explained mostly by the neutron deficiency that
is explained below\footnote{Note that in the nuclear network used in our study
no element heavier than germanium is included.}.

PISNe occur since the cores of their progenitors remain much less dense than
those of core-collapse SN progenitors \citep[c.f.][]{2012ARA&A..50..107L}.  
This makes neutronization during the final evolutionary stages much less significant than
in core-collapse progenitors. \citet{2002ApJ...567..532H} showed that this
results in a remarkable deficiency of odd-charged nuclei compared to even-charged
nuclei in the nucleosynthesis of Population~III PISNe.  Table~\ref{table:modeldata}
shows that our starting models have much larger neutron excesses
($\eta_\mathrm{i} \sim 10^{\,-4}$)  than the initial Population~III star models of
\citeauthor{2002ApJ...567..532H} ($\eta_\mathrm{i} \sim 10^{\,-7}$).  However, the
neutron excess in the central region ($\eta_\mathrm{c}$) where silicon burning
occurs does not increase much.  The maximum neutron excess achieved
at the center ($\eta^{\,\mathrm{max}}_\mathrm{c}$) in the 150~\Msun{} and
250~\Msun{} models is only $2 \times 10^{\,-4}$ and $1.6 \times 10^{\,-3}$,
respectively.  In the comparable Population~III helium star models of
\citet{2002ApJ...567..532H} (i.e. their 70~\Msun{} and 115~\Msun{} models; see
Figure~\ref{table:yields_selected}), the values of
$\eta^{\,\mathrm{max}}_\mathrm{c}$ are $2.8 \times 10^{\,-4}$ and $7.3 \times 10^{\,-4}$,
respectively (Table~\ref{table:modeldata}), which is a factor of 3$\ldots$5
lower.

This indicates that the neutron excess in PISNe
during the explosion phase does not increase by much more than about
$10^{\,-3}$ over the values at core helium exhaustion.  Interestingly, this
implies that even if the initial metallicity of a PISN progenitor was as high
as $Z_\odot${}, the neutron excess in the core would not become much
higher than about $10^{\,-3}$, which is the typical value of the neutron excess in
the innermost layers of solar metallicity massive stars at core helium
exhaustion.  This value is much smaller than the neutron excess achieved in the
silicon shell of a typical core-collapse progenitor ($\eta \cong 10^{\,-2}$),
where explosive nuclear burning occurs during the supernova explosion. 
We conclude that the odd-even effect is expected to be significant even in metal-rich PISNe.

\citet{2005IAUS..228..297H} argued that an unusually strong mixing of nitrogen
into the helium core of a PISN progenitor may make the odd-even effect as weak
as in the nucleosynthesis of a core-collapse supernova. However, to increase the
neutron excess to $\eta \cong 10^{\,-2}$ by such mixing is very
difficult to achieve in massive stars,  as discussed in
\citet{2012A&A...542A.113Y}.  It requires abundant production of primary
nitrogen in the first place, which in turn requires efficient chemical mixing
of carbon and oxygen into hydrogen shell burning during the post-main sequence
phases. Then, the primary nitrogen has to be mixed into the core of the star
to increase the neutron excess, via the $^{14}$N\,$(\alpha, \gamma)^{18}$F\,$(e^+,
\nu)^{18}$O reaction.  One possibility for such mixing is the penetration of the
helium-burning convective core into the hydrogen burning shell source, which is
often observed in massive Population~III star models \citep{2010ApJ...724..341H,
2012A&A...542A.113Y}.  Such mixing of nitrogen into the stellar core
by convection is not observed in our PISN progenitor models as shown in
Figure~\ref{figure:kipp}.  Even if it occurred, it would be difficult to 
enhance the neutron excess to more than about  $10^{\,-3}$
\citep{2012A&A...542A.113Y}.  Another possibility is rotationally
induced mixing. Our models are initially slow rotators and lost most of
their initial angular momentum via mass loss, rendering the role of rotation
unimportant.  \citet{2012A&A...542A.113Y} concluded that mixing of nitrogen
resulting from rotation may not enhance the neutron excess by more than about
$10^{\,-5}$ even for the extreme case of the so-called chemically homogeneous
evolution.  Furthermore, PISNe through chemically homogeneous evolution
are expected only at extremely low metallicity of $Z \lesssim 10^{\,-5}$
\citep{2006A&A...460..199Y, 2007A&A...475L..19L}.  We conclude
that the neutron excess does not exceed $\sim\,10^{\,-3}$ in most 
PISN progenitors. 
 
As expected from the above discussion, the odd-even effect in our models  also
appears to be strong (Figures~\ref{figure:prod1a}, \ref{figure:prod1b},
and~\ref{figure:prod2b}).  The ratio of the even- to odd-charged isotope mass
fractions reaches $10^{\,2}~-10^{\,5}$,  which is far from the observed values
in the solar system and in metal-poor stars
\citep{2004A&A...416.1117C,2004ApJ...603..708C,2005Natur.434..871F}.  However,
this effect is significantly weakened, compared to the case of the
corresponding Population~III star models,  for relatively light nuclei (i.e., lighter
than silicon for the 150~\Msun{} model and calcium for the 250~\Msun{} model
respectively).  This is because these elements are produced in the upper layers
of the star where neutronization during the explosive phase does not occur and
the degree of the odd-even effect is largely determined by the initial
metallicity of the star.  For example, the production factors of magnesium and
sodium differ by 1.2~dex in our 250~\Msun{} model, while this difference
increases to 2.1~dex in the corresponding Population~III model.

Note also that the overall production factors of our models are smaller than
those of the corresponding Population~III star models,  despite the fact that the
total yields of heavy elements are similar as shown in
Table~\ref{table:yields_selected}. For example, the production factors of iron
from our 250~\Msun{} model and a Population~III 115~\Msun{} helium star model are 59 and
125, respectively, while both models give the same total amount of iron 
(about 19~\Msun{}).  The reason for this difference is simply that the
presence of a hydrogen envelope is ignored in the case of the Population~III helium star models
(i.e., the helium core masses correspond to the initial masses), while in our
models the helium core masses are only  certain fractions of the initial
masses.

In the next section, we discuss the implications of this result for the
chemical evolution of galaxies.

\section{Implications for chemical evolution}
\label{sect:discussion1}

As mentioned above, very massive stars at relatively high metallicity are supposed to lose
too much mass to produce PISNe, and metal-poor environments are preferred for
PISN progenitors.  \citet{2007A&A...475L..19L} argued that the metallicity
threshold for PISNe ($Z_\mathrm{PISN}$) may be about $Z_\odot/10 \ldots
Z_\odot/3$. This conclusion is in an agreement with the recent result of
another detailed numerical study by \citet{2013MNRAS.433.1114Y}.  

Therefore, it is an important question how PISNe contributed to the chemical
evolution of galaxies in environments with $Z \lesssim Z_\mathrm{PISN}$.  
This critically depends on  how many PISN progenitors form as a function of time.
There are several possibilities for the formation mechanism of  very massive
stars \citep[cf.][]{2007ARA&A..45..481Z}, including very rapid mass accretion
\citep[e.g.][]{2009ApJ...703.1810H}, mergers in close binary systems and
stellar collisions \citep[e.g.,][]{2008A&A...477..223Y, 2012MNRAS.423.2203P}.  
Recent observations indicate that the upper stellar mass limit
($M_\mathrm{UP}$) may be as high as 180~\Msun{} in our Galaxy and 300~\Msun{}
in Large Magellanic Cloud~\citep{2010MNRAS.408..731C,2014ApJ...780..117S}.  Because of the paucity
of very massive stars discovered in the local Universe, these observations
still do not give a good constraint on the initial mass function for potential
progenitors of PISNe.  
If we simply assume the Salpeter-like IMF ($\Gamma = -2.3$), about 2\% of all
supernova progenitors ($M \gtrsim 12~M_\odot${}) have initial masses high
enough ($M \gtrsim 140~M_\odot${}) to produce a PISN.  

\begin{figure}[h]
\includegraphics[width=\columnwidth]{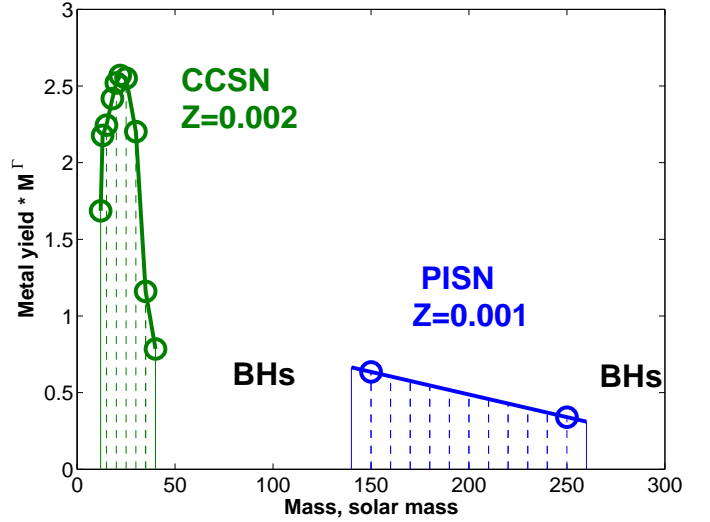}
\caption[The total metal yields of core-collapse SN models at $Z = 0.002$ and of PISNe at $Z = 0.001$ from the one generation of
stars.]
{The total metal yields of core-collapse SN models at $Z = 0.002$ 
provided by \citet{1995ApJS..101..181W} (the A-series SNe models; green line) and of PISNe at $Z = 0.001$ (blue line),
multiplied by the initial mass function probability ($\Phi(M) \propto M^{\,\Gamma}$\,), as a function of the initial mass.
The y-axis is given in arbitrary units.  Here, the PISN yields in the range 140~\Msun{}~--~260~\Msun{}     
are given by the extrapolation and interpolation of our 150~\Msun{} and 250~\Msun{} model results.  
We assumed a negligible metal yield for 40~$M_\odot < M < 140 M_\odot${} and above 260~\Msun{}.}
\label{figure:prod2a}
\end{figure}

To evaluate the contribution of PISNe to chemical evolution compared to that of core-collapse SNe, 
we calculated the production factor of major nuclei in the following way.  
The production factor integrated over an IMF ($\Phi(M) \propto M^{\,\Gamma}$) relative to solar abundances for a given isotope
is given by 
\begin{equation}
\label{equation:eq2}
P^\mathrm{int} = \frac{\int_{12}^{\,260} m_\mathrm{iso} \,\Phi(M)\,
dM}{\int_{12}^{\,260} X^{\,\odot}_\mathrm{iso} \, M \,\Phi(M) \,dM}
 = \frac{\int_{12}^{\,260} m_\mathrm{iso} \, M^{\,\Gamma} \,dM}
{\int_{12}^{\,260} X^{\,\odot}_\mathrm{iso} \, M^{\,\Gamma + 1} \,dM}~. 
\end{equation}
Here, the minimum and the maximum masses for supernova progenitors are assumed
to be 12~\Msun{} and 260~\Msun{}, respectively.  We adopt the core-collapse
SN~yields from \citet{1995ApJS..101..181W}.  
Since we have only two models
at 150~\Msun{} and 250~\Msun{}, we interpolate and extrapolate 
our results to cover the full PISN regime (140~--~260~\Msun) for this calculation, 
as shown in Figure~\ref{figure:prod2a}.  From the qualitative analysis of metal-free helium PISN models
\citep{2002ApJ...567..532H} we find that
linear interpolation gives about 20\% effect on weighted bulk yields, which correspond to 0.1~--~0.2~dex differences for final
bulk production factors.

This figure illustrates the contributions of core-collapse SNe and PISNe to the
chemical enrichment.  We assume that yields from core-collapse
SNe come from the explosions of massive stars in the mass range from 12~\Msun{}
to 40~\Msun{}.  These values are taken from the low energy explosion models
of massive stars at a metallicity of $Z = 0.002$ by \citet{1995ApJS..101..181W}.  
The integration over the hatched regions in the figure denotes
the IMF-weighted total amount of heavy elements (all elements
heavier than helium) ejected by the stars from one generation.  
Note that even though the number of stars in the PISN range is
significantly smaller than the number of core-collapse progenitors, the total amount of heavy elements
ejected from PISNe appears comparable to the integrated CCSN yield.  

Here we assume that stars with initial masses between 40~\Msun{} and
140~\Msun{}, and also above 260~\Msun{} do not considerably contribute to the
enrichment of surrounding medium with heavy elements.  
We should mention that massive stars lose mass through winds, 
which may be enhanced in metals (e.g. carbon) and they contribute to the galactic enrichment even though they form black holes
in the end.  However, the stellar winds, also those of the carbon-rich Wolf-Rayet stars, are reduced for lower initial iron
abundances \citep{2005A&A...442..587V}, such that their effect at the considered metallicities will be small.

\begin{figure}[h]
\includegraphics[width=\columnwidth]{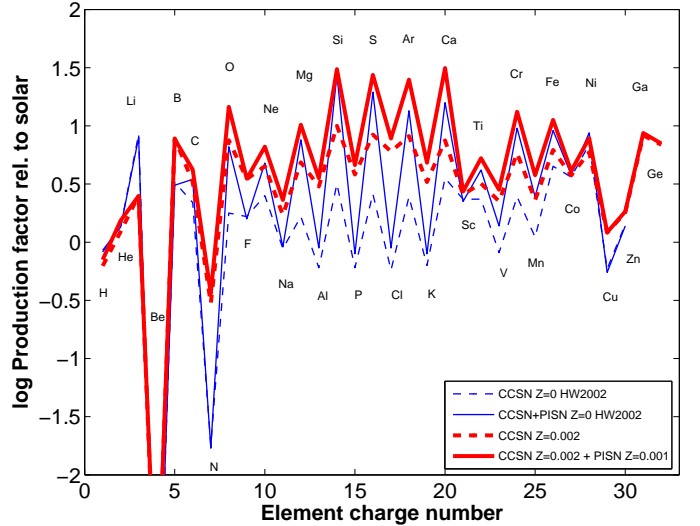}
\caption[Production factors of major elements from core-collapse SNe
and from both core-collapse and PISNe]
{Production factors relative to solar abundances of major elements from core-collapse SNe (dotted lines), 
and from both core-collapse and pair-instability SNe (solid lines).  
The blue thin lines are the results using the Population~III star models by \citet{1995ApJS..101..181W} and
\citet{2002ApJ...567..532H}.  The red thick lines correspond to the production factors using the values from
\citet{1995ApJS..101..181W} and present study (shown in Figure~\ref{figure:prod2a}). 
}\label{figure:prod2b}
\end{figure}

Figure~\ref{figure:prod2b} clearly indicates that the inclusion of PISN yields has
a strong impact on the total production factors even at finite metallicity: the
production factor of the even-charged nuclei is enhanced by a factor of 2~--~3
for most of the $\alpha-$elements with PISNe, while it is mostly negligible for
the odd-charged nuclei.   Note that this odd-even effect becomes strongest for
the elements between Al and Sc.  However, the odd-even effect is much weaker
in our models compared to the Population~III models. The Population~III
yields give almost 10~times higher production factors of even-charged nuclei
with the inclusion of PISNe.  This difference is mainly because the overall
core-collapse SN yields at a metallicity of $Z~=~0.002$ are significantly larger than those of
Population~III core-collapse SNe, and partly because our PISN models give a somewhat
weaker odd-even effect than the Population~III models as discussed above.

As discussed in Section~\ref{sect:intro1}, the event rate of PISNe is expected to decrease
to zero for metallicities higher than about $Z\,=\,Z_\odot/3${}.  The nucleosynthetic
signature of PISNe should be washed out by contribution of core-collapse SNe
as the metallicity reaches the solar value, and the effect of PISNe on
chemical evolution might not be found in Population~I stars.  However, from our study
we conclude that the impact of PISNe in the environment of $Z = 0.001 - 0.002$
may be still significant depending on the IMF, and should be tested in future
observations of Population~II stars with metallicities well below $Z = Z_\odot/3${}.

\section{Conclusions}
\label{sect:conclusion1}

In the frame of this study we calculated the evolution of two very massive
stellar models at a metallicity of $Z = 0.001$.  These two models have initial zero
age main sequence masses of 150~\Msun{} and 250~\Msun{}.  The models were evolved
through the core hydrogen, helium and carbon burning 
\citep{2007A&A...475L..19L} with the Binary Evolution Code {\sc{BEC}} of the
Bonn stellar physics group \citep{2006A&A...460..199Y}.  Here, we continued the
evolution using the same evolutionary code with an extended nuclear network
where 200~isotopes are considered.  We evolved these models through the
electron-positron pair creation phase and the consequent collapse and explosive oxygen and
silicon burning.  The 150~\Msun{} and 250~\Msun{} models eject a total amount of 64~\Msun{} and 111~\Msun{} of metals,
respectively.

We find that an excess production of even-charged elements compared to
odd-charged elements is still found in our models as in Population~III PISNe. However,
the odd-even effect is smaller for most of the $\alpha-$elements compared to
the case of Population~III stars because of the initially higher metallicity. The
nucleosynthetic pattern of the iron-group elements is critically determined by
the neutronization during the explosive burning and therefore less affected by
the initial metallicity. 

Given that our 150~\Msun{} and 250~\Msun{} models represent the low-mass and
high-mass ends of PISN regime respectively, this study allowed us to compare
the PISN nucleosynthesis with that of core-collapse supernovae at a similar
metallicity. We find that the impact of PISNe on the overall nucleosynthetic pattern is expected to be weaker at
$Z = 0.001 - 0.002$ than in the metal-free environment
(Figure~\ref{figure:prod2b}).  This is mainly because of the higher initial 
neutron excess of our models set by higher initial metallicity.

However, the total nucleosynthetic yields from both core-collapse SNe and PISNe
at $Z = 0.001 - 0.002$, assuming a Salpeter IMF, would result in the enhancement of
$\alpha-$elements by factors of 2~--~3 compared to the case without PISN
contribution (Figures~\ref{figure:prod2a} and \ref{figure:prod2b}).  This
analysis shows that PISNe at a metallicity of $Z = 0.001$ can contribute to the
enrichment of the interstellar medium with heavy elements in a similar way as
Population~III PISNe: the odd-even effect is still expected in metal-poor
stars at metallicities of about $0.001 - 0.002$, although
its degree would be reduced by factors of 3~--~4 compared to the prediction from Population~III
stars, as shown in Figure~\ref{figure:prod2b}.  Therefore, our models may be a
useful guide for interpreting future observations of the chemical abundances in Population~II stars
at $Z \approx 0.001 - 0.002$  to constrain the number of PISNe that might have
occurred in our Galaxy.

\begin{acknowledgements}
AK thanks Dr.~Sergey Blinnikov and Prof.~Dr.~Alexander Heger for fruitful and helpful discussions.  We also thank the referee for
useful comments which helped us to improve the draft.
\end{acknowledgements}

\bibliographystyle{aa}
\bibliography{references}

\input{append}

\end{document}

%% file: append.tex
\appendix
\section{Basic stellar structure equations}
\label{appendix:append}

The structure and evolution of stars are goverened by a set of partial differential equations.  The following are
{\emph{hydrodynamic}} equations which include inertia term.  However, they can be easily be converted into {\emph{hydrostatic}}
version while equating the inertia term with zero.

The set contains the equations of (1) continuity of mass,
(2) momentum and (3) energy.  Written in the vector form they are:
\begin{eqnarray}
\pd{\rho}{t} + \nabla\cdot(\rho\mb{u}) &=& 0 \\
\pd{\rho\mb{u}}{t} +\nabla\cdot(\mb{u}\otimes\rho\mb{u})
+\nabla p_g&=&0 \\
\pd{E}{t} +\nabla\cdot\left[\mb{u}\left(E+p_g\right)\right] &=&0 \,,
\label{eqn:euler}
\end{eqnarray}
where $\rho$ is the gas density, $\mb{u}$ its velocity, $p_g$ the gas
pressure, and $E$ the total gas energy per unit volume, $E = \rho
e_{\mr{int}}+\frac{1}{2} \rho u^2$, where $e_{\mr{int}}$ is the
internal energy per unit mass.  $\otimes$ denotes the outer product of
two vectors giving a $3\times3$ tensor.

In case of a non-rotating spherically symmetric star it is appropriate to use comoving
Lagrangian coordinates bound with the matter.  
Therefore, the above-written set of equations can be replaced by the following set:

\begin{eqnarray}
\left(\pd{r}{m}\right)_t & = & {1\over{4 \pi r^{\,2} \rho}}   \\
\left(\pd{r}{t}\right)_m & = & u                              \\
\left(\pd{P}{m}\right)_t & = & {G m\over{4 \pi r^{\,4}}} - {1\over{4 \pi r^{\,2}}} 
\left(\pd{u}{t}\right)_m \\
\left(\pd{l}{m}\right)_t & = & \varepsilon_{\mr{nuc}} + \varepsilon_{\mr{\nu}} + 
{P\over{\rho^{\,2}}} 
\left( \pd{\rho}{t} \right)_m - \left( \pd{\varepsilon}{t} \right)_m \\
\left(\pd{T}{m}\right)_t & = & - \nabla {G m T\over{4 \pi r^{\,4} P}} \left[1 + {r^{\,2}\over{G m}}
\left(\pd{u}{t}\right)_m \right] \,,
\label{eqn:lagrangian}
\end{eqnarray}
where $r$ is the radial distance of the shell to the centre of the star, 
$m$ is the mass contained -- serves as mass coordinate of the shell $m(r)=\int_{0}^{r} 4 \pi r^{\,2} \rho dr$, 
$\rho$ is the density in the shell, $u$ is the radial velocity, 
$P$ is the pressure, $G$ is the gravitational constant, 
$l$ is the local luminosity, $T$ is the temperature, 
$\varepsilon$ is the internal energy per unit mass,
$\varepsilon_{\mr{nuc}}$ corresponds to the energy release due to thermonuclear burning,
$\varepsilon_{\mr{\nu}}$ represents the local heat losses due to neutrino flux.

The details about this set of equations can be found, e.g., in \citet{1990sse..book.....K,1998PhDT........30H}.

The temperature gradient in radiative mass shells
\begin{equation}
\nabla \equiv \left(\pd{\ln T}{\ln P}\right)_t
\label{equation:grad}
\end{equation}
is given by radiative temperature gradient
\begin{equation}
\nabla_\mr{rad} = {3\over{16 \pi a c G}} {\varkappa l P\over{m T^{\,4}}} \left[1 + 
{r^{\,2}\over{G m}} \left(\pd{u}{t}\right)_m \right]^{\,-1} \,,
\label{equation:radgrad}
\end{equation}
where $a$ is the radiation constant, $c$ is the speed of light, and $\varkappa$ is the Rosseland mean opacity.  The opacities are
based on \citet{1994ApJ...437..879A} and \citet{1996ApJ...464..943I}.

The temperature gradient in convective shells is calculated using the mixing-length theory \citep[see
e.g.][and references therein]{1990sse..book.....K}.

The equation of state should be added to this set of equations, which binds the pressure $P$ with the temperature $T$.

The set of equations is the system of non-linear partial differential equations.  The evolution code solves it using the 
Newton-Raphson iteration method.  The
details about the numerical method can be found in \citet{1959ApJ...129..628H} and \citet{1964ApJ...139..306H}.